# Conventional superconductivity at 203 K at high pressures

A.P. Drozdov[1]*, M. I. Eremets[1]*, I. A. Troyan[1], V. Ksenofontov[2], S. I. Shylin[2]

[1] *Max-Planck Institut fur Chemie, Hahn-Meitner Weg 1,55128, Mainz, Germany*
[2] *Institute of Inorganic and Analytical Chemistry, Johannes Gutenberg-University Mainz, Staudingerweg 9, 55099 Mainz, Germany*

**A superconductor is a material that can conduct electricity with no resistance below its critical temperature ($T_c$). The highest $T_c$ that has been achieved in cuprates[1] is 133 K at ambient pressure[2] and 164 K at high pressures[3]. As the nature of superconductivity in these materials has still not been explained, the prospects for a higher $T_c$ are not clear. In contrast, the Bardeen-Cooper-Schrieffer (BCS) theory gives a guide for achieving high $T_c$ and does not put bounds on $T_c$, all that is needed is a favorable combination of high frequency phonons, strong electron-phonon coupling, and a high density of states. These conditions can be fulfilled for metallic hydrogen and covalent compounds dominated by hydrogen[4,5]. Numerous calculations support this idea and predict $T_c$ of 50-235 K for many hydrides[6] but only moderate $T_c$=17 K has been observed experimentally[7]. Here we studied sulfur hydride[8] where a $T_c\sim80$ K was predicted[9]. We found that it transforms to a metal at pressure ~90 GPa. With cooling superconductivity was found deduced from a sharp drop of the resistivity to zero and a decrease of $T_c$ with magnetic field. The pronounce isotope shift of $T_c$ in $D_2S$ is evidence of an electron-phonon mechanism of superconductivity that is consistent with the BCS scenario. The superconductivity has been confirmed by magnetic susceptibility measurements with $T_c$=203 K. The high $T_c$ superconductivity most likely is due to $H_3S$ which is formed from $H_2S$ under its decomposition under pressure. Even higher $T_c$, room temperature superconductivity, can be expected in other hydrogen–based materials since hydrogen atoms provide the high frequency phonon modes as well as the strong electron-phonon coupling.**

A search for high, room temperature conventional superconductivity is promising as the Bardeen−Cooper−Schrieffer (BCS) theory in the Eliashberg formulation puts no apparent limits on $T_c$. Materials with light elements are especially favorable as they provide high frequencies in the phonon spectrum. Indeed many superconductive materials have been found in this way, but only a moderately high $T_c$=39 K has been found in this search in $MgB_2$[10].

N. Ashcroft[4] turned attention to hydrogen which has very high vibrational frequencies due to the light hydrogen atom, and provides a strong electron-phonon interaction. Further calculations showed that metallic hydrogen should be a superconductor with a very high critical temperature $T_c$ ~100-240 K for molecular hydrogen, and $T_c$ = 300-350 K in the atomic phase at 500 GPa[11]. However superconductivity in pure hydrogen has not yet been found while the conductive and likely semimetallic state of hydrogen has been recently achieved[12]. Hydrogen dominated materials such as covalent hydrides $SiH_4$, $SnH_4$, etc. might also be good candidates for high $T_c$ superconductivity[5]. Similar to pure hydrogen, they have high Debye temperatures. Moreover, heavier elements might be beneficial as they contribute to the low frequencies that enhance electron phonon coupling. Importantly, lower pressures are required to metallize hydrides in comparison to pure hydrogen. Ashcroft's general idea was supported in numerous calculations[6,9] predicting high $T_c$'s for many hydrides. So far only low $T_c \sim 17$ K has been observed experimentally[7].

For the present study we selected $H_2S$ because it is relatively easy to handle, and predicted to transform to a metal and a superconductor at a low pressure of ~100 GPa with a high $T_c$ of ~80 K[9]. Experimentally $H_2S$ is known as a typical molecular compound with a rich phase diagram[13]. At about 96 GPa hydrogen sulfide transforms to a metal[14]. The transformation is complicated by the partial dissociation of $H_2S$ and the appearance of elemental sulfur at P>27 GPa at room temperature, and higher pressures at lower temperatures[13]. Therefore, the metallization of hydrogen sulfide can be explained by elemental sulfur which is known to become metallic above 95 GPa[15]. No experimental studies of hydrogen sulfide are known above 100 GPa.

In our typical experiment loading and pressure increasing were performed at temperatures of ~200 K which is essential for obtaining a good sample (see Methods summary). The Raman spectra of $H_2S$ and $D_2S$ were measured as the pressure increased and were in general agreement with the literature data[16,17] (see Extended Data Fig. 1). The sample starts to conduct at P~50 GPa. At this pressure it is a semiconductor as shown by the temperature dependence of the resistance and a pronounced photoconductivity. At 90-100 GPa the resistance drops further, and the temperature dependence becomes metallic. No photoconductive response is observed in this state. It is a poor metal – its resistivity at ~100 K is $\rho \approx 3 \times 10^{-5}$ Ohm m at 110 GPa and $\rho \approx 3 \times 10^{-7}$ Ohm m at ~200 GPa.

During the cooling of the metal at pressures of about 100 GPa (Fig. 1a) the resistance abruptly drops by three to four orders of magnitude indicating a transition to the superconducting state. At the next increase of pressure at low temperatures of T<100 K $T_c$ steadily increases with pressure. However at pressures of □160 GPa $T_c$ increases sharply (Fig. 1b). As higher temperatures of 150-250 K were involved in this pressure range we supposed that the increase of $T_c$ and the decrease of sample resistance during warming (Fig. 1a) can indicate a possible phase transformation with kinetics. Therefore in further experiments we anneal all samples by heating them to room or higher temperatures (Fig. 2a). This allowed us to obtain stable results, compare different isotopes, obtain the dependence of $T_c$ on pressure and magnetic field, and prove superconducting properties of the samples as follows.

(1) There is a sharp drop in resistivity with cooling indicating a phase transformation. The measured minimum resistance is at least as low ~$10^{-11}$ Ohm m – about two orders of magnitude less than for pure copper (Fig 1, see Extended Data Fig. 2e) measured at the same temperature[18].

(2) A strong isotope effect: $T_c$ shifts to lower temperatures for sulfur deuteride indicating phonon-assisted superconductivity (Fig. 2b,c). The BCS theory gives dependence of critical temperature on the atomic mass M as $T_c \sim M^{-\alpha}$ where $\alpha_H$ ~0.5. Comparison $T_c$'s in the pressure range P>170 GPa (Fig. 2c) gives $\alpha_H$ ~0.3.

(3) $T_c$ shifts to lower temperatures with available magnetic field up to 7 Tesla (Fig. 3). Much higher fields are required to destroy the superconductivity: extrapolation of $T_c(B)$ gives an estimate of a critical magnetic field as high as 70 T (Fig. 3).

(4) Finally, in magnetic susceptibility measurements (Fig. 4) a sharp transition from the diamagnetic to the paramagnetic state (Fig. 4a) was observed at the zero-field cooled (ZFC) curve. The onset temperature of the superconducting state $T_{onset} = 203(1)$ K, the width of the superconducting transition is nearly the same as in electrical measurements (Fig. 4a). Magnetization measurements M(H) at different temperatures (Fig. 4c) revealed a pronounced hysteresis indicating type II superconductivity. The magnetization decreases sharply at temperatures above 200 K indicating the onset of superconductivity at 203.5 K in agreement with the susceptibility measurements (Fig. 4a).

We have presented pure experimental evidence of superconductivity in sulfur hydride. However the compound responsible for the high $T_c$ is not obvious. The superconductivity measured in the low temperature runs (Fig. 1) likely relates to $H_2S$ as it is well described by calculations[9] for $H_2S$ – both the value of $T_c \sim 80$ K and its pressure behavior. However superconductivity with $T_c \sim 200$ K (Fig. 2) does not follow from these calculations. We suppose that it relates to the decomposition of $H_2S$ as high temperatures are required to reach the high $T_c$ (Fig. 2b). Precipitation of elemental sulfur upon the decomposition could be expected (which is well known at low pressures of P<100 GPa[13]), however superconducting transition in elemental sulfur occurs at significantly lower temperatures (Fig. 1b). Another expected product of decomposition of $H_2S$ is hydrogen. However the strong characteristic vibrational stretching mode from $H_2$ molecule was never observed in our Raman spectra as well as in Ref. [13]. Therefore we suppose that the dissociation of $H_2S$ is different with the creation of higher hydrides such as $3H_2S \rightarrow H_6S + 2S$ or $2H_2S \rightarrow H_4S + S$. It is natural to expect these reactions as the valency of sulfur is not only 2 but can also be 4, and 6. In fact calculations[9] indirectly support this hypothesis as the dissociation $H_2S \rightarrow H_2 + S$ was shown to be energetically very unfavorable. We found further theoretical support in Ref [19]. In this work the van der Waals compound[20] $(H_2S)_2H_2$ has been considered and shown that at pressures above 111 GPa $H_3S$ a molecular unit with $R3m$ structure is built, and above 180 GPa $H_6S$ units with $Im\text{-}3m$ structure are formed. The predicted $T_c$'s are respectively ~160 K and 190 K and the pressure dependences for these two phases are close to our experimental values (Fig. 2c). The phase transition between $R3m$ and $Im\text{-}3m$ structures is predicted at ~180 GPa. This likely relates to the changes in Raman spectra observed in the 170-180 GPa (see Extended Data Fig. 3). Our hypothesis on the transformation $H_2S$ to $H_6S$ ($H_3S$ stoichiometry) is fully supported by further calculations[21,22]. All the numerous works based on the $Im\text{-}3m$ structure[22-26] are consistent in prediction of $T_c > \sim 200$ K with the hydrogen sublattice giving the main contribution to superconductivity. Inclusion of zero point vibrations and anharmonicity in the calculations[23] corrected the calculated $T_c$ to ~190 K, and the isotope coefficient from $\alpha=0.5$ to $\alpha=0.35$ – both in agreement with the present work.

There is an interesting feature of the high $T_c$ $Im\text{-}3m$ phase: a strong covalent bonding between hydrogen and sulfur atoms[19]. This is in agreement with general assumption that material with high $T_c$ should have strong covalent bonding together with the high frequency modes in the phonon spectrum[27]. There are many materials which satisfy these conditions at ambient pressure: organics etc. but they are typical insulators. Principally, they can be tuned to a metallic state applying pressure or at ambient pressure by doping or gating.

**Methods Summary**

For electrical measurements we used diamond anvil cells (DAC) with anvils of the following shape: tip diameters of 200-300 μm bevelled at 7-8° to a culet of 40-80 μm. An insulating gasket is required to separate the metallic gasket from the electrodes. It was prepared in the following way (Extended Data Fig. 2). First a metallic gasket of T301 stainless steel (or Re) 250 μm thick was indented with about 17-20 GPa pressure. Then the bottom of the imprint of diameter ~200 μm was drilled out, and a powder insulating material was put in the imprint and pressed between the anvils to form a layer. The insulating layer was made of either Teflon, NaCl or $CaSO_4$ as these materials do not react with $H_2S$. The layer was pressed to obtain thickness in the center of ~3-5 μm to provide a stable clamping. Larger thickness leads to instability in the sample – it shifts or escapes under pressure, while with thinner gasket it is difficult to reach high pressures. A hole of diameter ~10-30 μm was then drilled in the insulating layer. Four Ti electrodes were sputtered on the diamond anvil. The electrodes were capped with Au to prevent the oxidation of the Ti. (To check a possible contribution of the diamond surface to the conductivity we prepared once a different configuration of electrodes: two electrodes were sputtered on one anvil and another two on another anvil similar to Ref. [12].) After preparation the electrodes the gasket was put back on the anvil and DAC was assembled so that separation between the anvils was about 20-100 μm (measured by interference fringes). The DAC was placed into a cryostat and cooled down to ~200 K (within the temperature range of liquid $H_2S$) and then $H_2S$ gas was put through a capillary into a rim around diamond anvil where it liquidified (Extended Data Fig. 4). $H_2S$ of 99,5% and $D_2S$ of 97% purity have been used. The filling was monitored visually (Extended Data Fig. 4,5) and the sample was identified by measuring Raman spectra. Then liquid $H_2S$ was clamped in the gasket hole by pushing the piston of the DAC with the aid of screws outside the cryostat. The thickness of the sample can be estimated to be few microns as measured from interference spectra through the clamped transparent sample. The thickness might be ~1 μm in if the sample expanded over the culet. After the clamping, DAC was heated to ~220 K to evaporate the rest of the $H_2S$, and then the pressure was further increased at this temperature. The pressures remained stable during the cooling within ±5 GPa. The low temperature loading seems to be required to prepare samples with high $T_c$. If $H_2S$ was loaded at room temperature in the gas loader, for example, only sulphur was detected in Raman and X-ray scattering. Apparently in this route the sample decomposes before reaching the required high pressure phase of $H_3S$. We did not explore all (P,T) paths to reach the state with high $T_c$. We found however that superconductivity is not observed in sample loaded at ~200 K but heated to room temperature at low pressure <~100 GPa.

The resistance and Raman spectra were measured during the pressurizing using the four-probe Van der Pauw method (Extended Data Fig. 2) with a current of 10 -10000 μA. The temperature was reliably determined by slow (warming rate of ~1 K/min) and allowing the DAC to equilibrate with attached thermometer. The determined $T_c$ was well reproduced in measurements with the PPMS and other setups. $T_c$ was determined as a point of the steepest change of resistance from the normal state (Fig. 2b).

The influence of the magnetic field on superconducting transitions has been measured with a nonmagnetic DAC (diameter 25 mm) in a PPMS6000 Quantum Design physical property device in a 4-300 K temperature range and fields up to 7 Tesla.

Magnetic susceptibility measurements were performed in MPMS Quantum Design setup. For that a miniature nonmagnetic cell made of Cu:Ti alloy working up to 200 GPa was designed (Fig. 4b). Samples of diameter of ~50-60 μm and thickness of few micrometers were prepared to provide a sufficient signal. Magnetic susceptibility measurements using a high-pressure cell were performed using a background subtraction feature of the MPMS software of SQUID magnetometer.

Pressure was determined by a diamond edge scale[28] at room and low temperatures. For optical measurements a Raman spectrometer was equipped with a nitrogen-cooled CCD and notch filters. The 632.8 nm line of a He-Ne laser was used to excite the Raman spectra and to determine pressure.

Acknowledgments. Support provided by the European Research Council under the 2010-Advanced Grant 267777 is gratefully acknowledged. We appreciate help provided in MPI Chemie by Prof. U.Pöschl. We are indebted to Patricia Alireza for genrous help with building of the new cell, as well as to J. Kamarad, S. Toser, and Jin for the help, K. Shimizu and his group for valuable cooperation, P. Chu and his group for many fruitful discussions and collaboration, L. Pietronero, M. Calandra and T.

# Figures

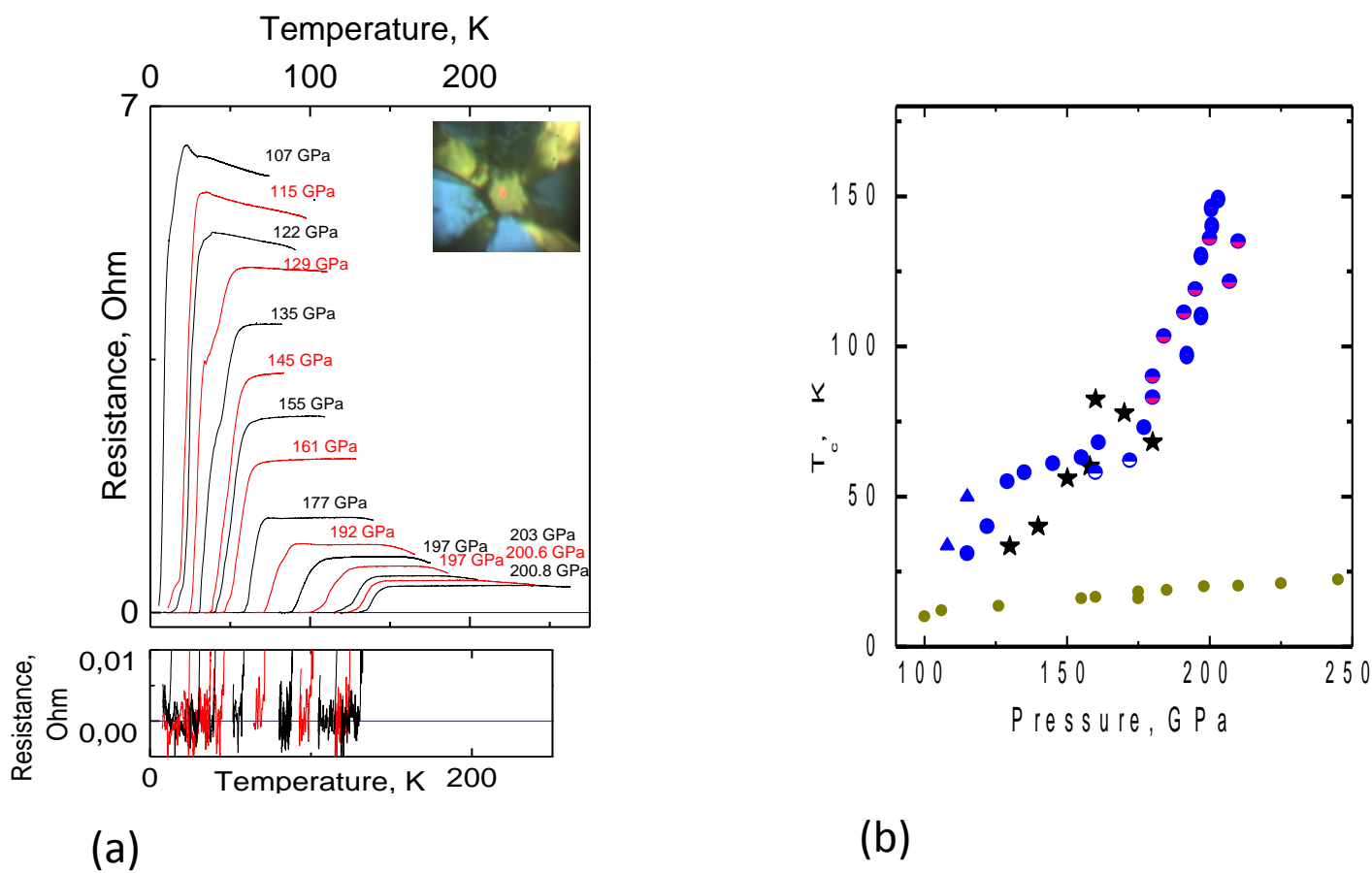

**Fig. 1. Temperature dependence of the resistance of sulfur hydride measured at different pressures.**

(a) Temperature dependence of resistance of sulfur hydride for different pressures. The pressure values are indicated near the corresponding plots. In this run the sample was clamped at T~220 K and the pressure was increased to ~100 GPa, and then the sample was cooled down. Plots at pressures <135 GPa were scaled (reduced 5-10 times) for easier comparison with the higher pressure steps. The resistance was measured with a current of 10 μA. Bottom: the resistance plots near zero. The resistance was measured with four electrodes deposited on a diamond anvil that touched the sample (inset). The diameters of the samples were~25 μm and the thickness ~1 μm.

(b) Blue round points represent $T_c$s determined from Fig. 1a. Other blue points were obtained in similar runs. Measurements at P>~160 GPa revealed a sharp increase of $T_c$. In this pressure range the R(T) measurements were performed over a larger temperature range up to 260 K, the corresponding experimental points are indicated by adding pink color. These points likely reflect a transient state for these particular pressure/temperature conditions. Further annealing of sample at room temperature would require stabilizing the sample (Fig. 2a). Black stars are calculations from Ref. 9. Dark yellow points are $T_c$'s of pure sulfur obtained with the same four-probe electrical measurement method. They are consistent with literature data[29] (susceptibility measurements) but have higher values at P>200 GPa.

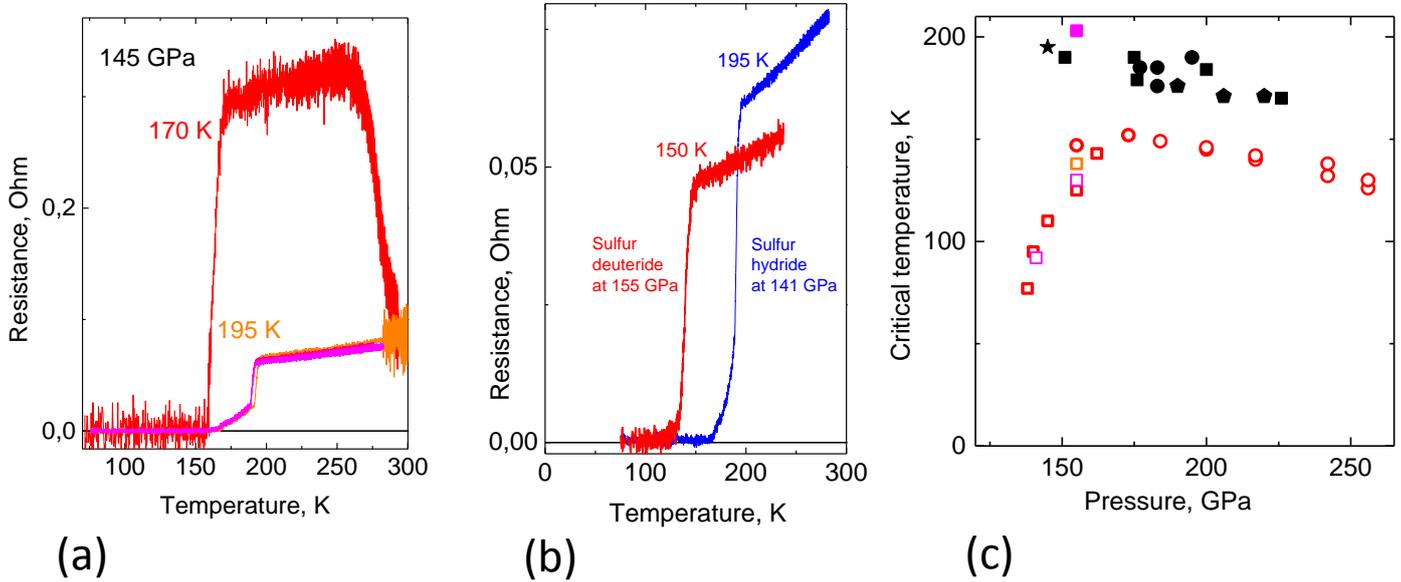

**Fig. 2. Pressure and temperature effects on T$_c$ in sulfur hydride and sulfur deuteride.**

(a) Changes of resistance and T$_c$ of sulfur hydride with temperature at constant pressure – annealing process. The sample was pressurized to 145 GPa at 220 K and then cooled to 100 K. During next slow warming ~1 K/min T$_c$ =170 K was determined. At temperatures above ~250 K the resistance dropped sharply, and at the next temperature run T$_c$ increased to ~195 K. This T$_c$ remained nearly the same for next two runs. (Note here that the only point for sulfur deuteride presented in arXiv:1412.0460 was determined without sample annealing, and T$_c$ would increase after annealing at room temperature.)

(b) (b) Typical superconductive steps for sulfur hydride and sulfur deuteride. The data are acquired during slow warming within several hours. The critical temperature T$_c$ is defined here as the sharp kink in the transition to normal metallic behavior. These curves were obtained after annealing at room temperature as shown in Fig. 2a.

(c) (c) Dependence of T$_c$ on pressure. The data on annealed samples are presented. Open color points refer to sulfur deuteride, and filled points to sulfur hydride. Magenta point was obtained in magnetic susceptibility measurements (Fig. 4a).

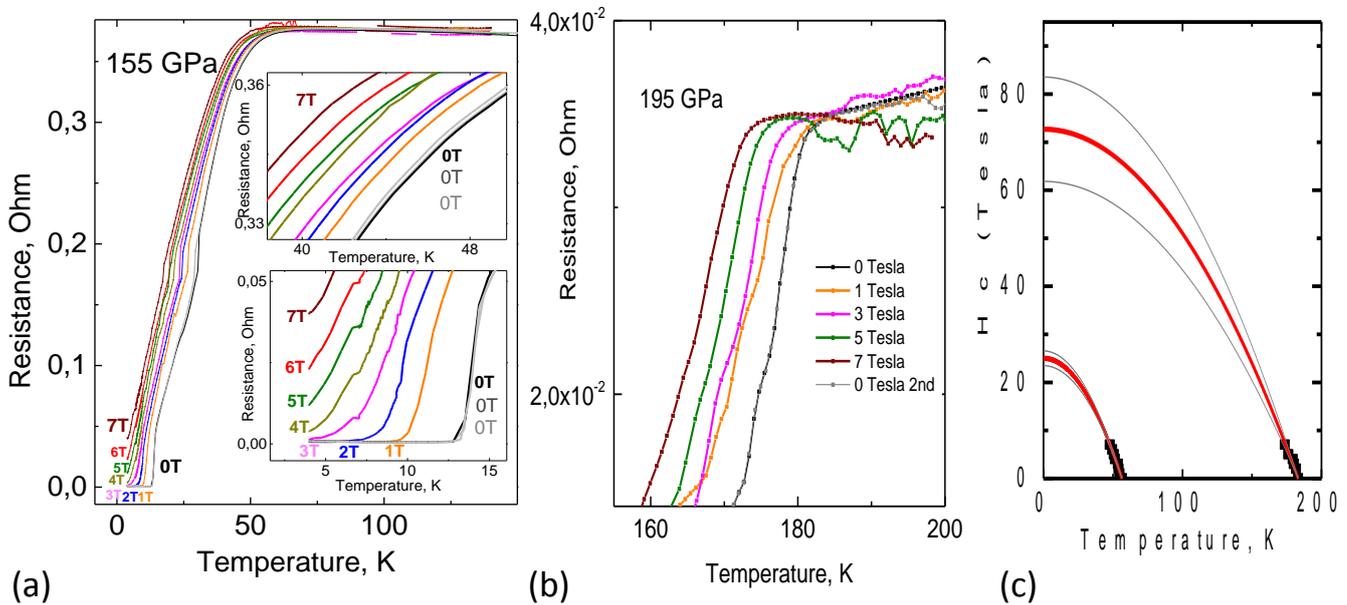

**Fig. 3. Temperature dependence of the resistance of sulfur hydride in different magnetic fields.**

(a) The shift of the ~60 K superconducting transition in the 0-7 T magnetic fields. The upper and low parts of the transition are shown enlarged in the insets. The temperature dependence of the resistance without magnetic field was measured three times: before applying the field, after applying 1,3,5,7 Tesla and finally after applying 2,4,6 Tesla (black, grey and dark grey colors).

(b) The same measurements but with the 185 K step.

(c) The shift of the critical temperature of the superconducting transition $T_c$ with magnetic field. To estimate the critical magnetic field $H_c$, the plots were extrapolated to high magnetic fields using $H_c(T) = H_{c0}(1-(T/T_c)^2)$ formula. The extrapolation has been done with 95% confidence (band shown as grey lines).

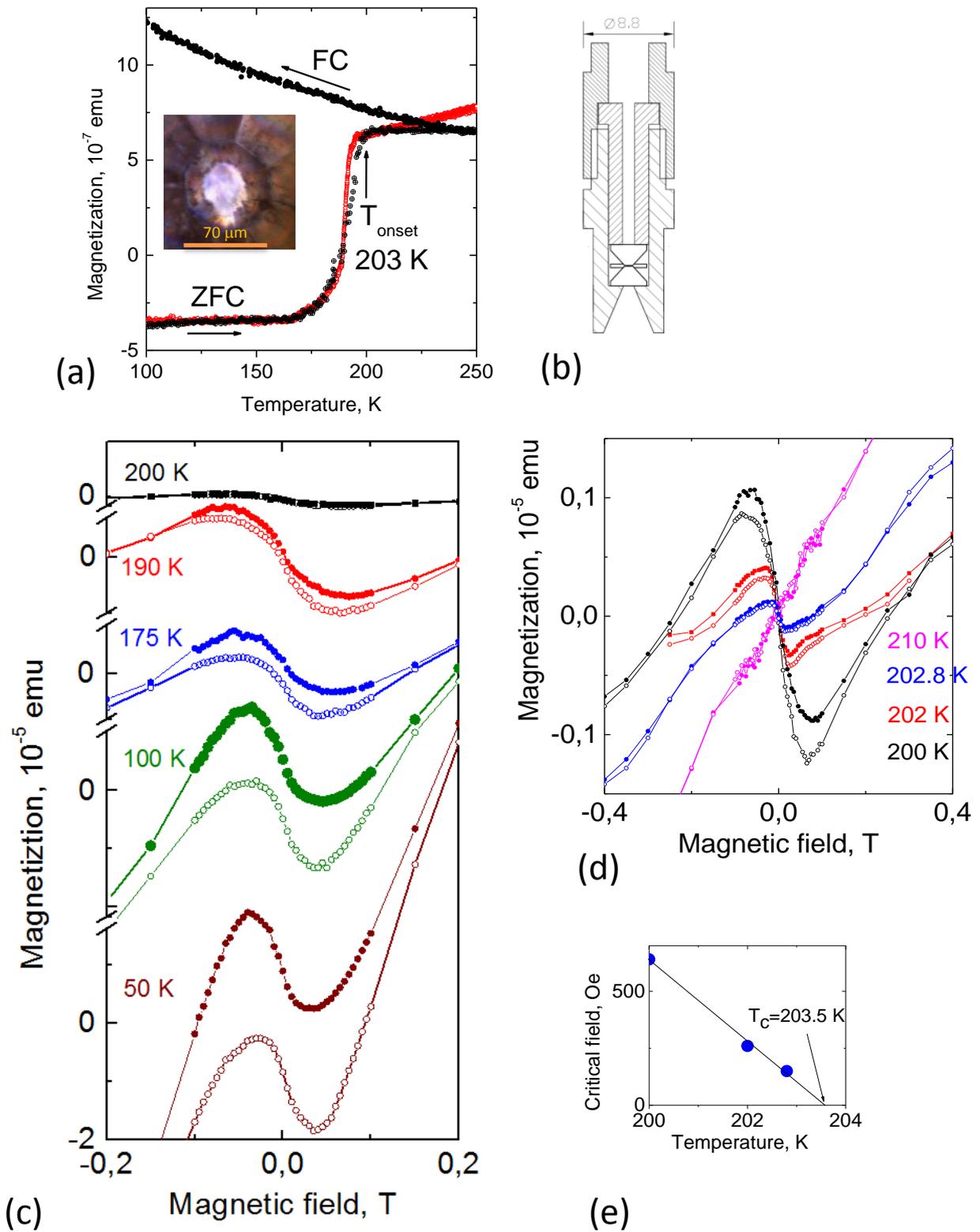

**Fig. 4. Magnetization measurements.**

(a) Temperature dependence of the magnetization of sulfur hydride at a pressure of 155 GPa in a the zero-field cooled (ZFC) and 20 Oe field cooled (FC) modes. The onset temperature is T$_{onset}$ = 203(1) K. For comparison, the superconducting step obtained for sulfur hydride from electrical measurements at 145 GPa is shown by red color. Resistivity data (T$_{onset}$ = 195 K) were scaled and settled vertically to compare with the magnetization data. The insert shows a sulfur hydride sample at 140 GPa in a Re gasket.

(b) Nonmagnetic diamond anvil cell.

(c) Magnetization measurements M(H) of sulfur hydride at a pressure of 155 GPa at different temperatures. The magnetization curves show hysteresis indicating the type II superconductor. The magnetization curves however are distorted by obvious paramagnetic input (observed in other superconductors too[30]). In our case the paramagnetic signal is likely from the DAC. Further study of the origin of this input is required. The paramagnetic background increases with a temperature decrease. The minima at the magnetization curves ($\sim$ 350 Oe) are result of the diamagnetic input from superconductivity and the paramagnetic background. The first critical field $H_{C1}\sim$100 Oe can be roughly estimated as a point where magnetization deviates from the linear behavior. At higher fields, magnetization increases due to the penetration of magnetic vortexes. As the sign of the field change reverses (full points), the magnetic flux in the Shubnikov phase remains trapped and therefore the back run is irreversible - returning branch of the magnetic cycle runs above the direct one. Hysteretic behavior of the magnetization becomes more clearly visible as the temperature decreases.

(d) At high temperatures T>200 K the magnetization decreases sharply.

(e) Extrapolation of the pronounced minima at the magnetization curves to higher temperatures gives the onset of superconductivity at T = 203.5 K.

**Extended Data for**

**Conventional superconductivity at  203 K at high pressures**

A.P. Drozdov[1]*, M. I. Eremets[1]*, I. A. Troyan[1], V. Ksenofontov[2], S. I. Shylin[2]

[1]*Max-Planck Institut fur Chemie, Hahn-Meitner Weg 1,55128, Mainz, Germany*

[2] *Institute of Inorganic and Analytical Chemistry, Johannes Gutenberg-University Mainz, Staudingerweg 9, 55099 Mainz, Germany*

corresponding author e-mail: m.eremets@mpic.de

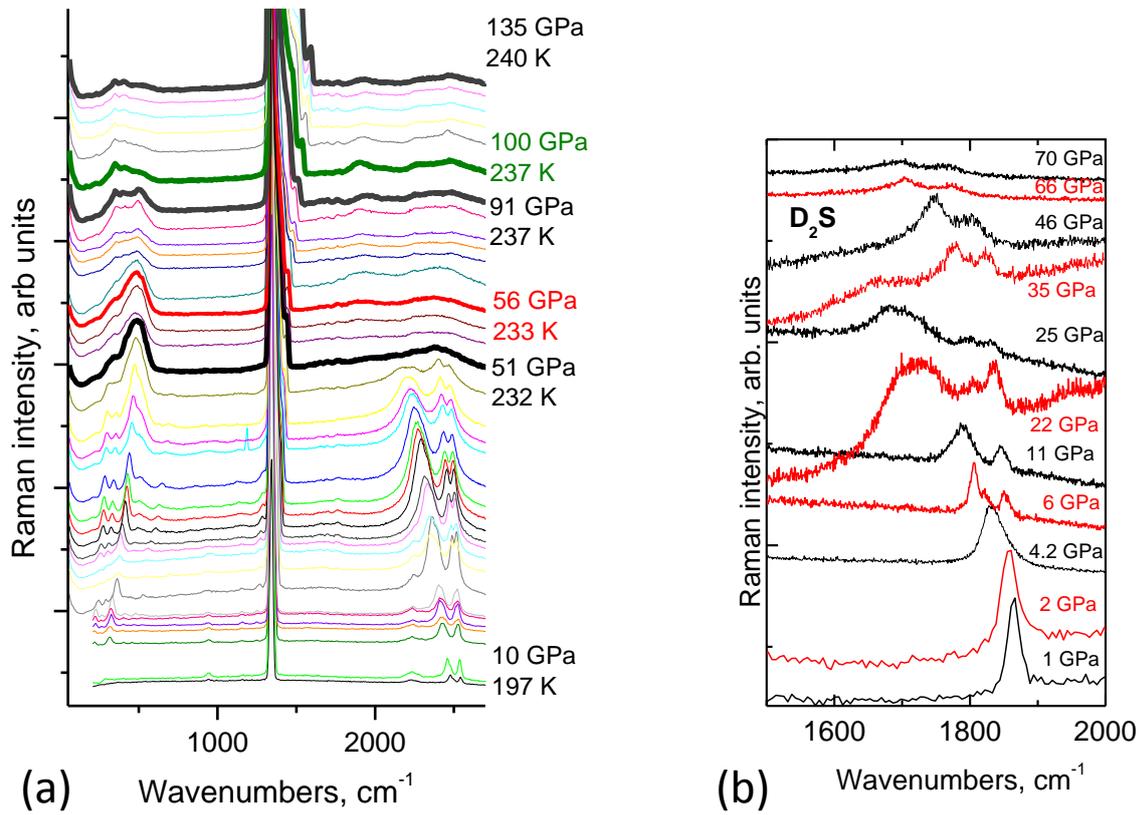

Fig. 1. Raman spectra of sulfur hydride at different pressures. (a) Spectra at increasing pressure at ~230 K. The spectra are shifted each other. The temperature of the measurement is also indicated. (b) Raman spectra of $D_2S$ measured at T~170 K.

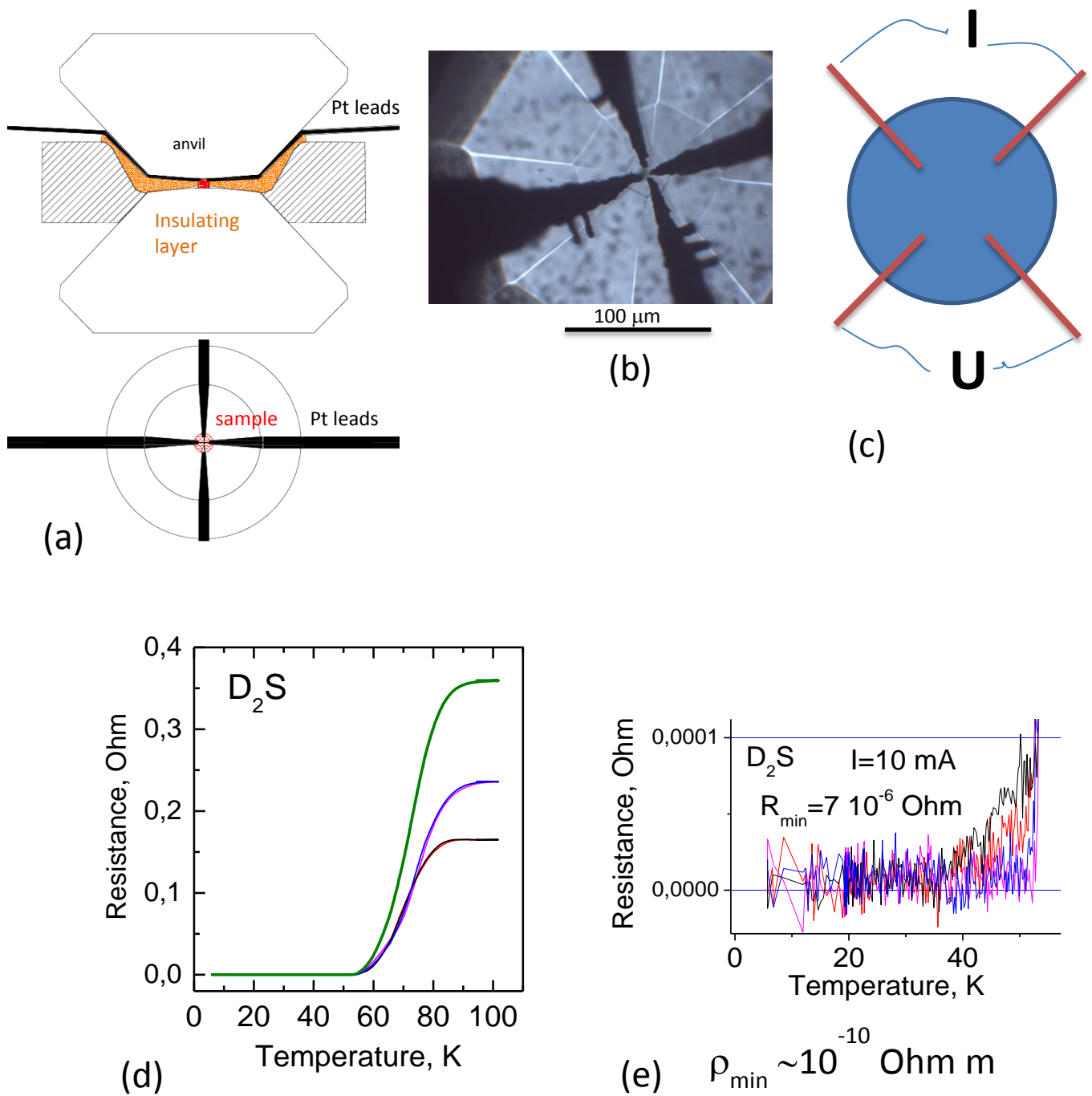

Fig. 2. Electrical measurements. (a) Schematic drawing of diamond anvils with electrical leads separated from metallic gasket with an insulating layer. (b) Ti electrodes sputtered on diamond anvil shown in transmitted light. (c) Scheme of the van der Pauw measurements. (d) Typical superconducting step measured in four channels (for different combinations of current and voltage leads shown in (c)). A sum resistance obtained from van der Pauw formula is shown by the green line. Note here that the SC transition was measured with the un-annealed sample (arXiv:1412.0460). After warming to room temperature and successive cooling the Tc should increase. (e) Residual resistance measured below the superconducting transition (d).

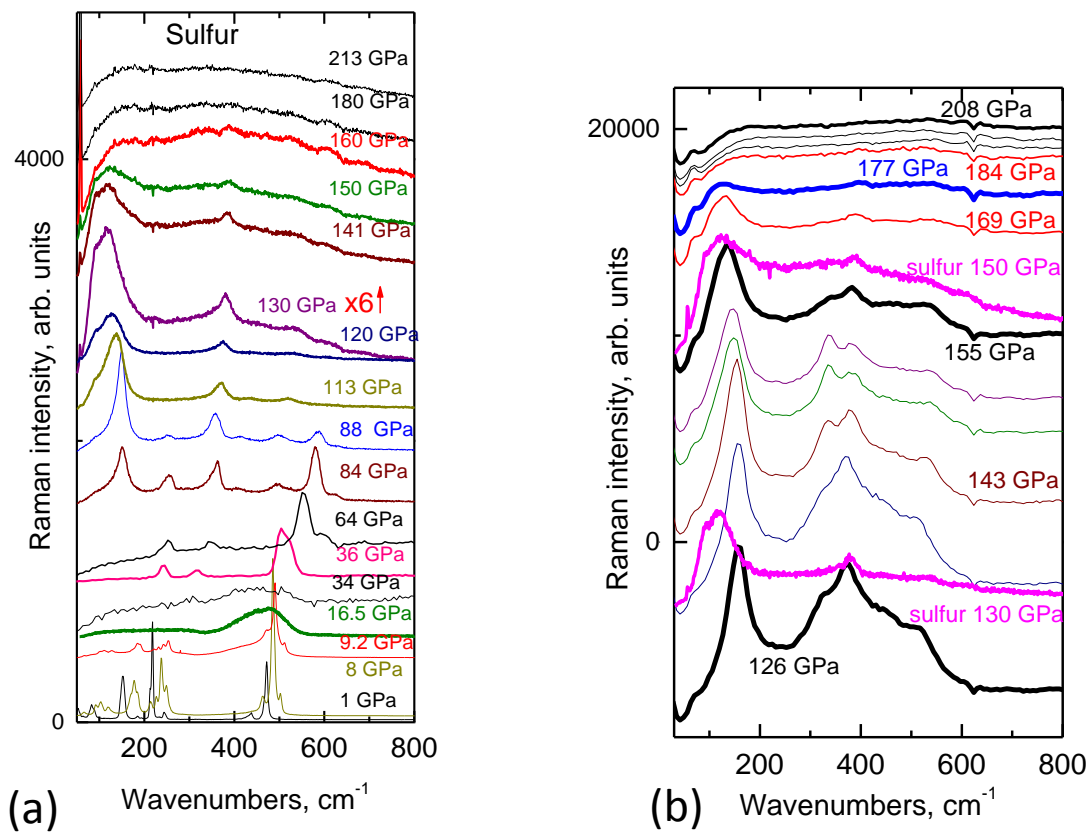

Fig. 3. Raman spectra of sulfur hydride compared with elemental sulfur. Ultralow luminescence synthetic diamond anvils allowed us to record the Raman spectra at high pressures in the metallic state.

Raman spectra of elemental sulfur at different pressures measured at room temperature over a wide pressure range. At low pressures the Raman signal is very strong but above ~10 GPa the intensity dramatically decreases, and a reflected signal appears. Above 80 GPa a new phase appeared which persists to 150 GPa in the metallic state. At 160 GPa the Raman spectrum disappears, likely because of transformation to the β-Po phase. The pressure of the transformation is in good agreement with our four probe electrical measurements of Tc (Fig. 2a). The electrical measurements in turn are good agreement with the susceptibility measurements (Gregoryanz et al., Phys. Rev. B 65, 064504) but we obtained noticeable higher Tc's at P> 200 GPa (to be published).

Raman spectra of sulfur hydride after the release of pressure from 208 GPa at room temperature. The spectra are much stronger than those from sulfur in the metallic state at high pressures. There is an apparent phase transition at ≈180 GPa.

There is an apparent difference in the Raman spectra of sulfur hydride (a) and sulfur (b): the peaks at ~100 cm$^{-1}$ shift relative to each other, the phase transformation are at different pressures: 160 GPa for sulfur and 180 GPa for sulfur hydride. Finally, the Raman spectra of sulfur hydride are significantly stronger.

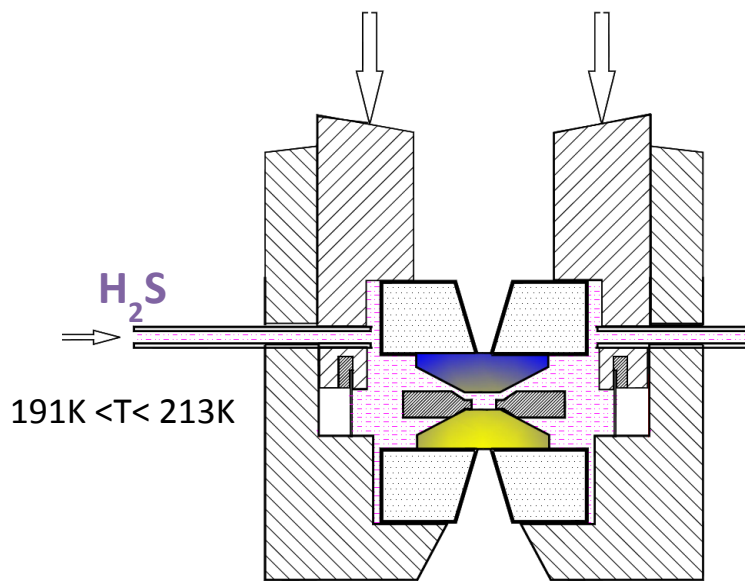

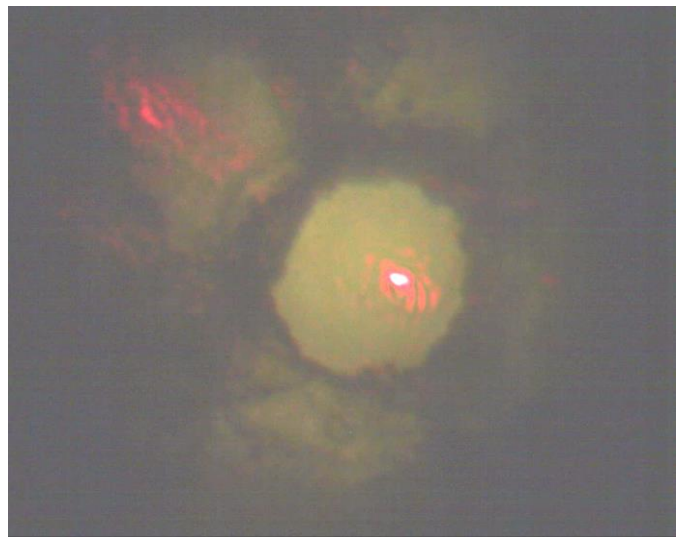

Fig. 4. Loading of H$_2$S. Gaseous H$_2$S is put through capillary to a rim around the diamond anvils. H$_2$S liquefies in the 191 K <T< 213 K temperature range and then it was clamped. The process of the loading is shown on the video (Video avaliable online at - https://vimeo.com/131914556). The hole in the transparent gasket (CaSO$_4$) is viewing through the diamond anvil. At T~200 K the gas cylinder with H$_2$S was opened and the gas condensed. At this moment the picture changed due to different refractive index of H$_2$S. After that second anvil with the sputtered electrodes was pushed, the hole was clamped. The sample changed color during the next application of pressure. Red point is from the focused HeNe laser beam.

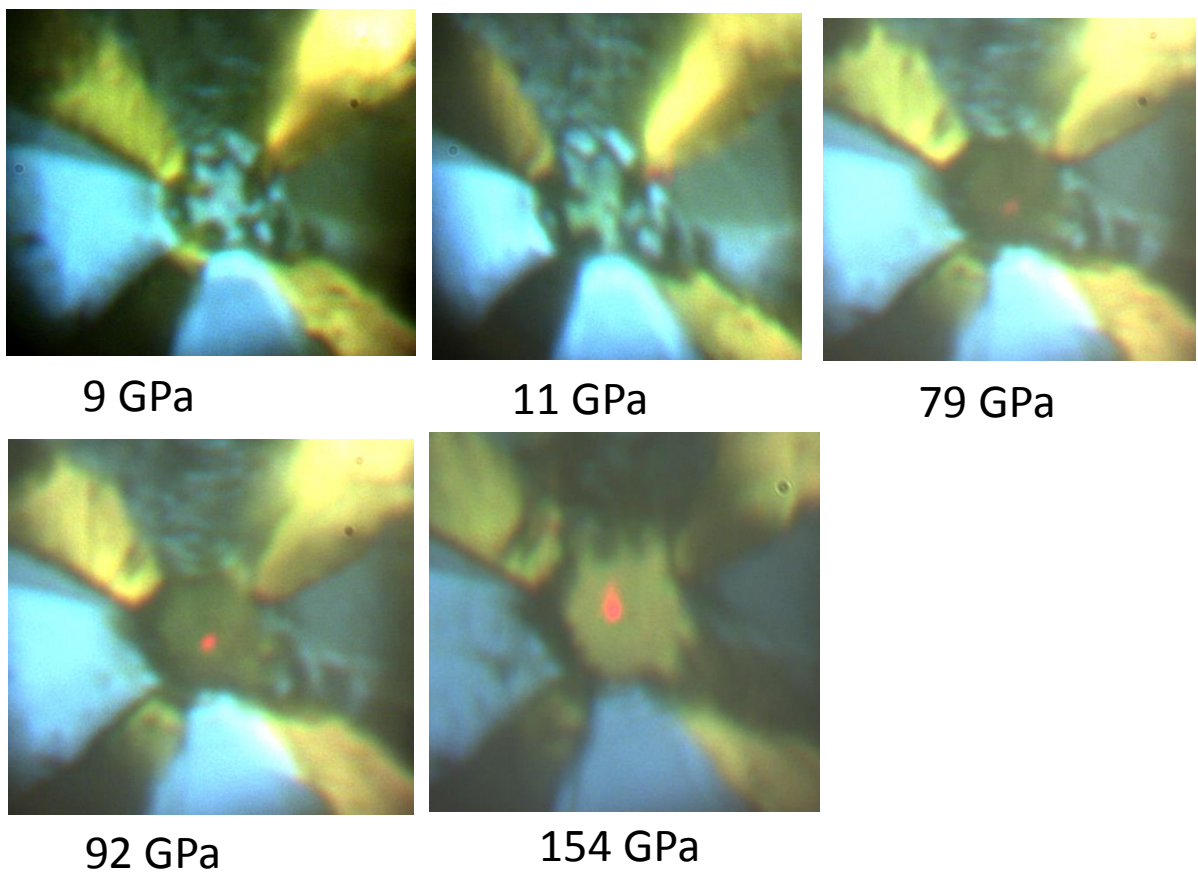

9 GPa          11 GPa          79 GPa

92 GPa          154 GPa

Fig. 5. View of D$_2$S sample with electrical leads and transparent gasket (CaSO$_4$) at different pressures. The photos were taken in a cryostat at 220 K with mixed transparent and reflection illumination.